\documentstyle[12pt,fleqn]{article}

\def\nc{N_{\rm c}}

\def\soc{{\rm C}_{60}}
\def\rug{{\rm C}_{70}}
\def\beeq{\begin{equation}}
\def\eneq{\end{equation}}
\def\beeqa{\begin{eqnarray}}
\def\eneqa{\end{eqnarray}}

\addtocounter{section}{-1}
\begin{document}

\begin{center}

{\large {\bf Polaron excitations in fullerenes:\\
Theory as $\pi$-conjugated systems}\\
(Review article in Prog. Theor. Phys.)\\
 }

\vspace{1cm}

{\rm Kikuo Harigaya}\\

\vspace{1cm}

{\sl Fundamental Physics Section, Physical Science Division,\\
Electrotechnical Laboratory,\\
Umezono 1-1-4, Tsukuba, Ibaraki 305}
\vspace{1cm}

(Received~~~~~~~~~~~~~~~~~~~~~~~~~~~~~~~~~~~)
\end{center}

\vspace{1cm}

\noindent
{\bf Abstract}\\
We review the recent theoretical treatment of fullerenes as $\pi$-conjugated
systems.  Polaronic properties due to the Jahn-Teller type effects
are mainly discussed. (1) A Su-Schrieffer-Heeger type electron-phonon
model is applied to fu\-lle\-re\-nes: C$_{60}$ and C$_{70}$, and is solved
with the adiabatic approximation to phonons.  When the system
(C$_{60}$ or C$_{70}$) is doped with one or two electrons (or holes),
the additional charges accumulate along almost an equatorial line
of the molecule.  The dimerization becomes the weakest along the
same line.  Two energy levels, the occupied state and the empty state,
intrude largely in the gap.  The intrusion is larger in C$_{70}$ than
in C$_{60}$.  These are ``polarons'' in doped fullerenes.  It is also found
that C$_{60}$ and C$_{70}$ are related mutually with respect to
electronical structures as well as lattice geometries.
(2) We apply the model to the fullerene epoxide C$_{60}$O.
It has the polaron-type lattice distortion around the
oxygen, and also shows the energy level intrusion in the gap.
(3) Optical properties of $\soc$ are calculated and discussed.
In the absorption of the doped molecule, a new peak structure
is present owing to the polaronic distortion.
In the luminescence of the neutral $\soc$,
the spacing between $H_g(8)$-phonon side-band
peaks and the relative intensities agree well with experiments.
In the dispersion of the third harmonic generation,
the magnitudes of $|\chi^{(3)}|$ agree with those of experiments
at the resonance of the lowest allowed transition as well as
in the region away from the resonance.

\pagebreak

%%%%%%%%%%%%%%%%%%%%%%%%%%  input document  %%%%%%%%%%%%%%%%%%%%%%%%%%%

\section{Introduction}

Recently, the ``fullerenes" C$_N$ which have the hollow cage
structures of carbons have been intensively investigated.
There are several experimental indications that the doped
fullerenes show polaronic properties due to the Jahn-Teller
distortion, for example: (1) The electron spin resonance
study$^{1)}$ on the radical anion of C$_{60}$ has
revealed the small $g$-factor, $g=1.9991$, and this is
associated with the residual orbital angular momentum due
to the Jahn-Teller distortion. (2) Photoemission studies$^{2)}$
of C$_{60}$ and C$_{70}$ doped with alkali metals have shown
peak structures, which cannot be described by a simple
band-filling picture. (3) When poly(3-alkylthiophene) is
doped with $\soc$,$^{3)}$ interband absorption of
the polymer is remarkably suppressed and
the new absorption peak evolves in the low energy range.
The Jahn-Teller splitting of LUMO in C$_{60}^-$ state and/or
the Coulomb attraction of positively charged polaron to
C$_{60}^-$ might occur. (4) The luminescence of
neutral $\soc$ has been measured.$^{4)}$ There are two peaks
around 1.5 and 1.7eV below the gap energy 1.9eV, interpreted
by the effect of the polaron exciton. In addition, the experiments
on the dynamics of photoexcited states have
shown the interesting roles of polarons.$^{5)}$

In this article, we first review the recent investigation of
polaronic excitations in the C$_{60}$ and $\rug$ molecules,
and discuss lattice distortion and reconstruction of
electronic levels upon doping.  We have described C$_{60}$
and $\rug$ as an electron-phonon system and have extended
the Su-Schrieffer-Heeger (SSH) model$^{6)}$ of conjugated polymers.
We have calculated for systems where one or two electrons
are added or removed.  We shall discuss properties of
``polarons" in fullerenes, which have been reported
in detail in refs. 7-9.

We have found that sites, where additional charges are prone to
accumulate, are common to $\soc$ and $\rug$.  They are along
the equatorial line in $\soc$. Ten more carbons are inserted
between these sites in $\rug$.  In this regard, we should bear
in mind that $\rug$ is made from $\soc$, by division into two parts
and adding ten carbons.  Thus, there are relations of electronical
properties as well as the structural relation between
(doped as well as undoped) $\soc$ and $\rug$.

Next, we review the application of the extended SSH model to the
fullerene epoxide $\soc$O.$^{10)}$  The dimerization
has been found to become weaker around the sites near the
oxygen.  Two energy levels intrude largely in the gap. These
polaronic features are certainly the effects of the external
potential given by the additional oxygen.

Finally, we look at optical properties of $\soc$.  We
have considered optical absorption spectra of the
doped molecules,$^{9)}$ luminescence from the photo-excited
neutral $\soc$,$^{11)}$ and the dispersion of the third harmonic
generation (THG).$^{12)}$  In the absorption of the doped $\soc$,
a new peak structure is present owing to the polaronic distortion.
In the luminescence, the spacing between $H_g(8)$-phonon side-band
peaks and the relative intensities agree well with experiments.
In the dispersion of the THG, the magnitudes of $|\chi^{(3)}|$
agree with those of experiments at the resonance of the lowest
allowed transition as well as in the region away from the resonance.

This article is organized as follows.  In \S 2, the model is presented.
Polarons in $\soc$ and $\rug$ are discussed in the following two sections.
In \S 5, the extended SSH system is applied to the fullerene epoxide.
In \S 6, we show the optical properties.  We close this article
with brief remarks in \S 7.

\section{Model}

We use the extended SSH hamiltonian for the topological geometries
of fullerenes: $\soc$ and $\rug$.  The model is:
\beeq
H = \sum_{\langle i,j \rangle, \sigma} ( - t_0 + \alpha y_{i,j} )
( c_{i,\sigma}^\dagger c_{j,\sigma} + {\rm h.c.} )
+ \frac{K}{2} \sum_{\langle i,j \rangle} y_{i,j}^2,
\eneq
where $c_{i,\sigma}$ is an annhilation operator of a $\pi$-electron; the
quantity $t_0$ is the hopping integral of the ideal undimerized system;
$\alpha$ is the electron-phonon coupling; $y_{i,j}$ indicates the bond
variable which measures the length change of the bond between
the $i$- and $j$-th sites from that of the undimerized system;
the sum is taken over nearest neighbor pairs $\langle i j \rangle$;
the second term is the elastic energy of the lattice; and the
quantity $K$ is the spring constant.  This model is solved with
the assumption of the adiabatic approximation and by an iteration method.

\section{Polarons in C$_{\rm 60}$}

We have taken $t_0 = 2.5$eV which has been used in the two-dimensional
graphite plane$^{13)}$ and polyacetylene.$^{6)}$
Two quantities, $\alpha = 6.31$eV/\AA{\ } and
$K=49.7$eV/\AA$^2$, have been determined so that the length
difference between the short and long bonds in $\soc$ is the
experimentally observed value: 0.05\AA.$^{14)}$  Here, the
dimensionless electron phonon coupling $\lambda
\equiv 2\alpha^2 / \pi K t_0$ has been taken as 0.2 as in
polyacetylene.$^{6)}$  The number of electrons $N_{\rm el}$
has been varied within $-2 \leq N_{\rm c} \leq 2$, where
$N_{\rm c} = N_{\rm el} - N$, and $N$ is the number of carbon atoms.

%Fig. 1
First, we discuss lattice and electronic structures of
$\soc$.$^{7,9)}$  The lattice configurations of the doped
systems are shown in Fig. 1(a) and (b).  We show three kinds
of the shorter bonds.  The shortest bonds, d, are represented
by the thick lines.  The second shortest ones, b, are shown
by the usual double lines.  The dashed lines indicate the
third shortest bonds.  They are the bonds f in Fig. 1(a) and
bonds g in Fig. 1(b). Other longer bonds are not shown.  The
figures are the same for the electron and hole dopings.
When the change in the number of electrons is one, the change
in the electron density is the largest at the sites at the ends
of dashed lines, namely, points D.  The dashed lines are mostly
located along an equatorial line of C$_{60}$.  The absolute value of
the length of the bonds g is the smallest of the four kinds
of bonds with negative bond variables.  The dimerization becomes
the weakest along this equatorial line.  The distortion of the
lattice is similar to that of a polaron$^{15)}$ in conjugated
polymers.  When the change in the electron number is two,
configurations of dashed lines along the equatorial line change,
as shown in Fig. 1(b). The ordering of bonds, f and g,
with respect to the bond variable is reversed. Other configurations
are the same. The change in the electron density is also the largest
at points D.  Therefore, polaronic distortion persists when
the doping proceeds from one to two electrons (or holes).

%Fig. 2
Next, we look at changes in the electronic level structures.
They are shown in Fig. 2.  When the system is doped,
the degeneracy decreases due to the reduced symmetry.
This reduction comes from the deformation of the lattice.
This is one of the Jahn-Teller distortions.  The removal of
the degeneracies of energy levels is due to the $H_g$
distortion.$^{16)}$  When $|N_{\rm c}| = 1$ and 2, the
highest level, which splits from the highest occupied molecular
orbital (HOMO) of the neutral
system, is nondegenerate.  Its energy shifts upward.  In contrast,
the other two levels shift only slightly.  Similarly, the
lowest unoccupied molecular orbital (LUMO)
of the neutral system splits into two levels.  The energy of
the nondegenerate level shifts downward, while change of
the energy of the doubly degenerate level is small.  This change
in the level structures is common to two cases of the electron
and hole dopings.  The change is similar to that in the polaron
formation$^{15)}$ in conjugated polymers.

\section{Polarons in C$_{\rm 70}$}

%Fig. 1
We describe changes in lattice structures and electron distributions
of the doped $\rug$.$^{8,9)}$  The dimerization strengths change
their values mostly along the ring-like part
shown in Fig. 1(c), while the patterns with the mirror reflection
symmetry persist.  Change in electron density at sites E is
very small.  This is a consequence of the fact that dimerization
almost disappears along bonds f and g. The property of the part
along the equatorial line is similar to that of the graphite plane.
The strengths of the dimerization change largely along bonds,
from a to e, upon doping.  The additional charges tend to
accumulate near these bonds.  The positions D, where the additional
charges accumulate most densely, correspond to the sites D of
$\soc$.   When we make $\rug$ from $\soc$, sites E are added in the
interval, but the property, that additional charges tend to
accumulate at sites D, persists for $\rug$.  This finding is
quite interesting.

%Fig. 3
We discuss structures of electronic energy levels.  We show the
changes in electronic structures in Fig. 3.  Two levels have
already intruded in the gap in the neutral $\rug$.  This is
due to the structural elongation from $\soc$ to $\rug$.
When the system is doped with up to two electrons or holes,
the HOMO and LUMO of the neutral system largely extend into the gap.
The positions of the other levels change only slightly.
The magnitude of level intrusion is larger than that in
$\soc$ due to nondegenerate levels near the gap at $\nc = 0$.
The HOMO and LUMO of the neutral system have large amplitude
at sites, from A to D.  The amplitude at D is the largest.
The amplitude at E is very small. Therefore, the additional
charge is prone to accumulate most at sites D.

\section{Fullerene Epoxide}

The fullerene epoxide $\soc$O shown in Fig. 1(d) has been produced
by a large amount.$^{17)}$  In this section, we look at the
dimerization and energy level structures.  Details are discussed
in ref. 10.  The oxygen atom is treated by an impurity model
with one effective atomic level.  The following term is added to
the extended SSH model:
\beeq
H_{\rm O} = E_{\rm O} \sum_s d_s^\dagger d_s
- t_{\rm O} \sum_s [ d_s^\dagger (c_{1,s} + c_{2,s}) + {\rm h.c.} ].
\eneq
The carbons bonding to the oxygen are numbered as 1 and 2.

%Fig. 4
We find the following properties:  The di\-merization becomes
weaker around the oxygen as shown in Fig. 1(d).  Two localized
states appear deep in the gap.  Figure 4 shows the data for
$t_{\rm O} = 0.5 t_0$.  The HOMO and LUMO are shown by full
and open circles, respectively.  The next HOMO (NHOMO) and
next LUMO (NLUMO) are represented by the small squares
which are connected by curves.  Other energy levels are not
shown.  All the levels are nondegenerate. When E$_{\rm O}$ is
varied, the energies of NHOMO and NLUMO do not change so much as
those of HOMO and LUMO.  This is the consequence of the fact
that the wave functions of NHOMO and NLUMO spread almost
over the $\soc$ while those of HOMO and LUMO are localized
around the defect.  Thus, we can regard the HOMO and LUMO
as impurity states which are well known in bulk semiconductors.
Therefore, we conclude that the energy gap of $\soc$ itself
is less affected by the defect while two new localized states
are emitted into the gap due to the defect potential.  This
property is like that of polarons but mainly due to the
external potential from the oxygen.  We note that optical transition
between them is allowed.  This accords with the recent optical
absorption data.$^{18)}$

\section{Optical Properties of C$_{\rm 60}$}

For optical properties, we consider optical absorption spectra of the
doped molecules (ref. 9), luminescence from the photoexcited
neutral $\soc$ (ref. 11), and the dispersion of the THG (ref. 12).

%Fig. 5
We show how the ``polarons'' in $\soc$ would be observed in optical
absorption.$^{9)}$ Figure 5 shows the results of undoped and
electron-doped systems with $0 \leq \nc \leq 2$.  We only show
data of the electron-doped cases.  Figure 5(a) is the data
of the undoped system.  There are two peaks in the figure.
The peak at 2.9eV is the transition between the HOMO and the NLUMO.
The other peak at 3.1eV is the transition between the NHOMO
and the LUMO.  The transition between the HOMO and the LUMO
is forbidden and does not appear in the figure.  Figures 5(b)
and (c) show the data of the systems with $\nc = 1$ and 2,
respectively.  The two large peaks in Fig. 5(a) now have small
substructures due to the level splittings.  In addition, there
appears a new peak at low energy ($\sim$0.7eV).    This peak
corresponds to the transition between the singly occupied
molecular orbital and the NLUMO
etc., when $\nc = 1$.  It corresponds to the transition
between the LUMO and the NLUMO
etc., when $\nc = 2$.  Therefore, the new peak at low energy is
due to the splitting of the LUMO.

%Fig. 6
Next, we consider quantum lattice fluctuations and
discuss the phonon side bands in the luminescence spectrum.$^{11)}$
To calculate the luminescence, we use a collective coordinate
method$^{19)}$ which resembles the displacement of the carbon
atoms of the $H_g$(8) phonon mode and extrapolates between the
ground state ``dimerization" and the exciton polaron.  Wave
functions of the singly occupied molecular orbitals of the exciton
polaron have large amplitudes at twenty sites along the equatorial line.
The coordinate describing the weakening of the
dimerization along the equator is assigned to the twenty carbons.
The collective coordinate Schr\"{o}dinger equation is solved
and the luminescence is calculated by the formula used by
Friedman and Su.$^{19)}$  It is assumed that disorders and/or
solid state effects make the dipole-forbidden transition allowed
partially.$^{4)}$  The result is shown by plots in Fig. 6.
The curves are the envelopes of the experimental data.$^{4,20)}$
There is good agreement for the existing luminescence
peak spacing.  This indicates that the $H_g$(8) mode frequency
1575cm$^{-1}$ is quantitatively derived by the coordinate.
We also find fair agreement for the relative intensity.
The difficulty in making the $\soc$ thin films and different
experimental conditions would be the origins of the variety of
experimental data.  Anyway, our success has revealed the
importance of the intramolecular electron-phonon couplings in $\soc$.

%Fig. 7
Finally, we look at spectral dispersions of the THG of the
neutral $\soc$.$^{12)}$  We display the absolute value in Fig. 7.
In the bottom of the figure, we show the energies of the dipole
allowed excitations, where three-photon resonances can appear,
and also the energies of the forbidden excitations multiplied by $3/2$,
where two-photon resonances can appear.  The peaks in the THG spectrum
can be assigned as two- or three-photon resonances.   We point out
three properties: (1) The magnitude of $\chi^{(3)}$ at
$\omega = 0$ is $1.22 \times 10^{-12}$esu and is similar to
the magnitudes in the THG experiments: $4 \times 10^{-12}$esu
at $3\omega \simeq 1.6$eV (ref. 21) and $7 \times 10^{-12}$esu
at $3\omega \simeq 3.6$eV (refs. 21 and 22). Here, we compare
the magnitudes at frequencies far from the resonances.
(2) The value of the THG around the peak at 2.5eV is
of the order of $10^{-11}$esu.  This well explains the magnitude
$2.7 \times 10^{-11}$esu at the peak centered around 2.8eV (ref. 21).
A larger broadening in our theory would yield better agreement
with the experiment.  We note that several authors have pointed out
the similar property that the third harmonic generation of $\soc$
can be explained by the free electron theory$^{23,24)}$ or the
calculation within the Hartree-Fock approximation.$^{25)}$
(3) The three-photon peaks, at $3\omega \simeq 6.1$ and 6.3eV,
have remarkably large strengths. This large enhancement would
be due to the fact that there are two-photon resonant $3\omega
\simeq 6.1$, 6.4, and 6.5eV, meaning that double resonance enhancement
occurs.  Electron correlations might change this third consequence.

\section{Concluding Remarks}

We have reviewed the recent theoretical treatment of the fullerenes
as the $\pi$-conjugated systems.  Even though the extended SSH model
is very simple, we have derived interesting general properties
about polaronic lattice distortions and energy level intrusions.
There have been several origins of the polaronic changes:
the doping of additional charges to $\soc$ and $\rug$
(refs. 7 and 8), the structural elongation from $\soc$ to $\rug$ (ref. 9),
and effects of the external potential in $\soc$O (ref. 10).

Recently, several interesting optical experiments have been
reported. The properties more or less resemble those of
the conjugated polymers, which have been explained by the
one-dimensional models like the SSH hamiltonian.$^{15)}$
We have derived quantitative agreements with the luminescence
and THG experiments.  Thus, the theoretical model of
fullerenes as the $\pi$-conjugated systems is a powerful method
for investigating the dynamical as well as static properties.

We have neglected various effects: Coulomb interactions among
$\pi$-electrons, thermal fluctuations of phonons, interactions
between molecules, and so on.  They should be taken into
account when we compare the theory with various experiments$^{1-5)}$
consistently.  These effects would be important when the excitation
energy is large in the dynamical processes.  They pose interesting
problems for future works.

{}~~~~~~~

\noindent
{\bf Acknowledgements}\\
Useful discussion with Prof. Y. Wada, Dr. K. Yamaji,
Prof. H. Fukuyama, Prof. W. P. Su, Dr. A. Oshiyama, Dr. S. Saito,
Dr. N. Hamada, Prof. G. A. Gehring, Dr. M. Fujita,
Dr. Y. Asai, Dr. A. Terai, Dr. T. Yanagisawa, and Dr. Y. Shimoi
is acknowledged.  Fruitful collaboration with Prof. B. Friedman
and Dr. S. Abe is also acknowledged.
Numerical calculations have been performed
on FACOM M-780/20 and M-1800/30 of the Research Information
Processing System, Agency of Industrial Science and Technology, Japan.

\pagebreak

\begin{flushleft}
{\bf References}
\end{flushleft}

\noindent
1) T. Kato, T. Kodama, M. Oyama, S. O\-ka\-za\-ki, T. Shida, T. Nakagawa,
Y. Matsui, S. Suzuki, H. Shiromaru, K. Yamauchi and Y. Achiba,
Chem. Phys. Lett. {\bf 180} (1991), 446.\\
2) T. Takahashi, S. Suzuki, T. Morikawa,  H. Katayama-Yoshida,
S. Hasegawa, H. Inokuchi, K. Seki, K. Kikuchi, S. Suzuki, K. Ikemoto
and Y. Achiba, Phys. Rev. Lett. {\bf 68} (1992), 1232;
C. T. Chen, L H. Tjeng, P. Rudolf, G. Meigs, L. E. Rowe, J. Chen, J. P.
McCauley Jr., A. B. Smith III, A. R. McGhie, W. J. Romanow
and E. W. Plummer, Nature {\bf 352} (1991), 603.\\
3) S. Morita, A. A. Zakhidov and K. Yoshino, Solid State Commun. {\bf 82}
(1992), 249; S. Morita, A. A. Zakhidov, T. Kawai, H. Araki and K. Yoshino,
Jpn. J. Appl. Phys. {\bf 31} (1992), L890.\\
4) M. Matus, H. Kuzmany and E. Sohmen, Phys. Rev. Lett. {\bf 68}
(1992), 2822.\\
5) P. A. Lane, L. S. Swanson, Q. X. Ni, J. Shinar, J. P. Engel, T. J. Barton
and L. Jones, Phys. Rev. Lett. {\bf 68} (1992), 887.\\
6) W. P. Su, J. R. Schrieffer and A. J. Heeger, Phys. Rev. B {\bf 22}
(1980), 2099.\\
7) K. Harigaya, J. Phys. Soc. Jpn. {\bf 60} (1991), 4001;
B. Friedman, Phys. Rev. B {\bf 45} (1992), 1454.\\
8) K. Harigaya, Chem. Phys. Lett. {\bf 189} (1992), 79.\\
9) K. Harigaya, Phys. Rev. B {\bf 45} (1992), 13676.\\
10) K. Harigaya, J. Phys.: Condens. Matter {\bf 4} (1992), 6769.\\
11) B. Friedman and K. Harigaya, (preprint).\\
12) K. Harigaya and S. Abe, Jpn. J. Appl. Phys. {\bf 31} (1992), L887.\\
13) G. W. Hayden and E. J. Mele, Phys. Rev. B {\bf 36} (1987), 5010.\\
14) C. S. Yannoni, P. P. Bernier, D. S. Bethune,
G. Meijer and J. R. Salem, J. Am. Chem. Soc. {\bf 113} (1991), 3190.\\
15) A. J. Heeger, S. Kivelson, J. R. Schrieffer and W. P. Su,
Rev. Mod. Phys. {\bf 60} (1988), 781.\\
16) C. M. Varma, J. Zaanen and K. Raghavachari, Science
{\bf 254} (1991), 989.\\
17) J. M. Wood, B. Kahr, S. H. Hoke II, L. Dejarme, R. G. Cooks
and D. Ben-Amotz, J. Am.  Chem. Soc. {\bf 113} (1991), 5907.\\
18) K. M. Creegan, J. L. Robbins, W. K. Robbins, J. M. Millar, R. D. Sherwood,
P. J. Tindall and D. M. Cox, J. Am. Chem. Soc. {\bf 114} (1992), 1103.\\
19) B. Friedman and W. P. Su, Phys. Rev. B {\bf 39} (1989), 5152.\\
20) C. Reber, L. Yee, J. McKiernan, J. I. Zink, R. S. Williams,
W. M. Tong, D. A. A. Ohlberg, R. L. Whetten and F. Diederich,
J. Phys. Chem. {\bf 95} (1991), 2127.\\
21) J. S. Meth, H. Vanherzeele and Y. Wang, (preprint).\\
22) Z. H. Kafafi, J. R. Lindle, R. G. S. Pong, F. J. Bartoli,
L. J. Lingg and J. Milliken, Chem. Phys. Lett. {\bf 188} (1992), 492.\\
23) A. Ros\'{e}n and E. Westin, (preprint).\\
24) S. V. Nair and K. C. Rustagi, (preprint).\\
25) Z. Shuai and J. L. Br\'{e}das, (preprint).\\

\pagebreak

\begin{flushleft}
{\bf Figure Captions}
\end{flushleft}

\noindent
Fig. 1.  Lattice structures of doped $\soc$ [(a) $| \nc |=1$ and (b)
$| \nc | = 2$], (c) neutral $\rug$, and (d) fullerene epoxide $\soc$O.

{}~

\noindent
Fig. 2.  Energy-level structures of $\soc$ with $-2 \leq N_{\rm c} \leq 2$.
The line length is proportional to the degeneracy of the energy level.
The shortest line is for the nondegenerate level.  The arrow shows the
position of the Fermi level.

{}~

\noindent
Fig. 3.  Energy-level structures of $\rug$ with $-2 \leq N_{\rm c} \leq 2$.
The notations are the same as in Fig. 2.

{}~

\noindent
Fig. 4.  Energy levels of $\soc$O as a function of $E_{\rm O}$.
We use $t_{\rm O} = 0.5t_0$.

{}~

\noindent
Fig. 5. Optical absorption of $\soc$ for (a) $\nc=0$, (b) $\nc=1$, and
(c) $\nc=2$.

{}~

\noindent
Fig. 6. Relative intensity of the luminescence from the photoexcited $\soc$.
The large dots are the calculated data.  The solid curve is the
envelope of the experimental data by Matus et al. (ref. 4)
and the dashed curve is the data by Reber et al. (ref. 20).

{}~

\noindent
Fig. 7.  Dispersion of third harmonic generation of the neutral $\soc$.
The closed circles are the data by Meth et al. (ref. 21).

{}~

\noindent
{\bf Note: Figures will be sent by the conventional mail.  Please
request to the e-mail address: harigaya@etl.go.jp.}

%%%%%%%%%%%%%%%%%%%%%%%%%%%%%%%%%%%%%%%%%%%%%%%%%%%%%%%%%%%%%%%%%%%%%%%

\end{document}